\documentclass[twocolumn,showpacs,preprintnumbers,amsmath,amssymb,aps]{revtex4}

\usepackage{graphicx}
\usepackage{dcolumn}
\usepackage{subfig}

\begin{document}

\title{Detection of Single Ion Spectra by Coulomb Crystal Heating}

\author{Craig R. Clark}
\author{James E. Goeders}
\author{Yatis K. Dodia}
\author{C. Ricardo Viteri}
\author{Kenneth R. Brown}
\altaffiliation{Electronic mail: ken.brown@chemistry.gatech.edu}
\affiliation{Schools of Chemistry and Biochemistry; Computational Science and Engineering; and Physics, Georgia Institute of Technology, Atlanta, Georgia 30332, USA}

\date{\today}

\begin{abstract}
The coupled motion of ions in a radiofrequency trap has been used to connect the frequency-dependent laser-induced heating of a sympathetically cooled spectroscopy ion with changes in the fluorescence of a laser-cooled control ion. This technique, sympathetic heating spectroscopy, is demonstrated using two isotopes of calcium.  In the experiment, a few scattered photons from the spectroscopy ion are transformed into a large deviation from the steady-state fluorescence of the control ion.  This allows us to detect an optical transition where the number of scattered photons is below our fluorescence detection limit. Possible applications of the technique to molecular ion spectroscopy are briefly discussed.
\end{abstract}


\maketitle

\section{Introduction}
Laser-cooled ions in linear Paul traps are an ideal tool for studying gas phase atomic and molecular ions at very low temperatures. The large trapping depth (on the order of eV) and the wide mass acceptance range allow the simultaneous trapping of different ion species. An ion that can be laser cooled can be used to sympathetically cool another species through the Coulombic interaction between the ions~\cite{BabaJJAP1996, MolhavePRA2000, DrewsenPRL2004, RothPRL2005, BlythePRL2005, RyjkovPRA2006, OstendorfPRL2006, OffenbergPRA2008}. Sympathetic cooling brings the motion of all trapped ions to the  equilibrium temperature of the laser-cooled ions~\cite{Alekseev1995} in a time proportional to the secular frequency of the trap~\cite{Papenbrock2002, Baba2002}. Controlled ensembles of cold atomic and molecular ions have the potential for many applications, including: quantum information processing~\cite{BarrettPRA2003, SchmidtScience2005}, ultra-high-resolution spectroscopy~\cite{SchmidtScience2005, BatteigerPRA2009, WolfPRL2009, WolfPRA2008V78, HerrmannPRL2009}, optical clocks~\cite{RosenbandPRL2007, Rosenband2008}, nano deposition of dopant atoms in semiconductors~\cite{SchnitzlerPRL2009}, and studies of molecular properties and chemical reactions~\cite{WillitschPRL2008, BellMP2009, CarrNJP2009, BellFD2009, OkadaJPB2003, DrewsenIJMS2003, RothPRA2006V73, RothJPB2006, StaanumPRL2008, HojbjerrePRA2008, OffenbergJPB2009}.


At low temperatures the trapped ions form an ordered structure known as a Coulomb crystal. This presents an opportunity to heat the crystal with one ion and detect that heating with another. Laser-cooled fluorescence mass spectrometry (LCFMS)~\cite{BabaJJAP1996} uses a probe voltage at the secular frequency of the target ion to heat the crystal. The fluorescence drops as the cooling ions are sympathetically heated and Doppler shifted with respect to the cooling laser. The charge-to-mass ratio of the target ion can then be determined. This method has been applied to detect the photofragmentation of molecular ions in a Coulomb crystal~\cite{ KoelemeijPRL2007, HojbjerrePRA2008, OffenbergJPB2009,HojbjerreNJP2009}.

Resonance-enhanced multiphoton dissociation, in combination with LCFMS, can be used to gain high-resolution spectral information as demonstrated with a crystal of HD$^{+}$ and Be$^+$ ions~\cite{KoelemeijPRL2007}. First, a vibrational overtone line was excited in HD$^+$ with an infrared diode laser locked to a stable frequency comb. An UV laser then transferred the excited population to a dissociative state. The HD$^+$ population decay can then be monitored by observing the Be$^+$ fluorescence. The resulting spectral line had a width of 40 MHz, dominated by Doppler shifts due to the micromotion (driven motion) inherent in large crystals in a radiofrequency ion trap. Recently, the same rotational state selective dissociation spectroscopy technique has been utilized to map the state populations of translationally and vibrationally cold molecular ions and achieve a high degree of rotational cooling~\cite{HojbjerreNJP2009, Staanum2010, Schneider2010}.

Narrower linewidths can be achieved by limiting the ions to a linear chain.  Quantum logic spectroscopy (QLS)  transfers information between two trapped ions, a logic ion and a spectroscopy ion, through the quantized vibrational motion of the crystal. The logic ion serves as a quantum sensor for detecting transitions in the spectroscopy ion~\cite{SchmidtScience2005}. The reported absolute frequency measurements on single trapped ions using QLS have been performed on narrow transitions~\cite{RosenbandPRL2007, Rosenband2008} where the secular frequency of the ions in the trap exceeded the transition linewidth. In this strong-binding limit~\cite{WinelandPRA2009}, the absorption spectrum consists of a carrier and a number of motional sidebands separated by the secular frequency. Initialization in QLS experiments requires cooling of the vibrational modes of the crystal to the ground state, which is achieved by addressing the motional sidebands~\cite{Monroe1995}.

There are a wide variety of interesting transitions that cannot be studied with QLS due to their large linewidths. High-precision spectroscopy outside the strong-binding limit is challenging since the spectroscopy laser induces detuning-dependent heating and cooling which distorts the line profile.  Recently, experiments in the weak binding limit have been performed using low-intensity spectroscopy laser beams on a chain of sympathetically cooled ions~\cite{WolfPRA2008V78, HerrmannPRL2009, BatteigerPRA2009}. The sympathetic cooling removes the heating limitation by counteracting the back action of the interrogating lasers. High resolution is achieved by calibrating the probe laser to an absolute frequency by referencing it to a frequency comb or using the comb light directly ~\cite{WolfPRL2009}. In this case, the ion fluorescence is measured directly as the constantly sympathetically cooled ion absorbs the minimal heating due to the low intensity lasers.

In this work, the frequency-dependent heating of a spectroscopy ion is measured by observing the fluorescence of a second ion (control ion) as the system is recooled. We refer to this method as sympathetic heating spectroscopy (SHS). The method is demonstrated on two isotopes of calcium: $^{40}$Ca$^{+}$, the control ion,  and $^{44}$Ca$^{+}$, the spectroscopy ion.  Even a low scattering rate of photons from the spectroscopy ion can create a significant stochastic optical force that builds up quickly with the laser interaction time ($t_{{\rm heat}}$), and dramatically changes the trajectory of both ions. This results in a large Doppler shift of the control ion which can be observed in the recooling process. Laser induced fluorescence (LIF) experiments using similar very low laser intensities will require long photon counting times to acquire line profiles with decent signal to noise. Potentially, SHS can become an effective tool to study dipole transitions that are weak or fall in regions of the electromagnetic spectrum where the sensitivity of detectors is marginal or non-existent.

\section{Experimental}

\subsection{Experimental Setup}

The experiments are performed in a linear Paul trap held in vacuum at 1$\times 10^{-10}$ torr. The trap is a five-segment version of the three-segment trap described in Ref.~\cite{FurukawaJJAP2005} and a duplicate of the trap used in Ref.~\cite{LeibrandtPRL2009}. The trap is driven at 14.5 MHz and the secular frequencies for $^{40}$Ca$^{+}$ are measured to be 0.5, 1.0, and 1.3 MHz.  The ion micromotion is minimized by applying compensation voltages while measuring the correlation between the fluorescence and the trap drive~\cite{Berkeland1998}. The Doppler recooling method~\cite{EpsteinPRA2007,WesenbergPRA2007} was used to measure the trap heating. For a single ion, no heating was observed for dark times up to 20 s. 

The experiment requires the loading of one $^{40}$Ca$^+$ and  one $^{44}$Ca$^+$. This is accomplished using resonance-enhanced two-photon ionization~\cite{KjaergaardAPB2000,GuldeAPB2001,LucasPRA2004} and a natural Ca source (97\% $^{40}$Ca, 2\% $^{44}$Ca). The neutral calcium is isotopically selected using the 4s$^{2}$ $^{1}$S$_{0}\leftrightarrow$ 4s5p$^{1}$P$_{1}$ transition at $423$ nm~\cite{LucasPRA2004,TanakaAPB2005}. A second photon at 377 nm then ionizes the excited calcium.

\begin{figure}[htp]
\includegraphics[scale=0.35]{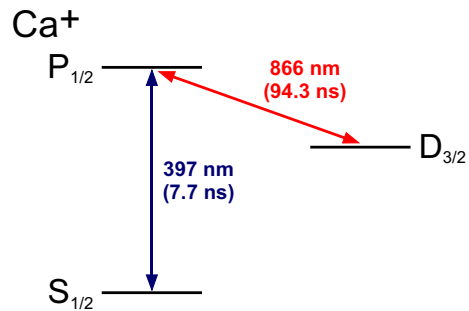}
\caption{\small (Color online) Energy levels with corresponding transition wavelengths and lifetimes~\cite{James1998} used for Doppler cooling of $^{40}$Ca$^{+}$ and laser-induced heating of $^{44}$Ca$^{+}$. The relative isotope shifts are $842$ MHz and $-4.5$ GHz for the 397-nm and the 866-nm transitions, respectively~\cite{LucasPRA2004}.}
\label{fig:EnergyDiagram}
\end{figure}

\indent The Doppler cooling of  Ca$^{+}$ requires two lasers: the main cooling laser at 397 nm, which addresses the S$_{1/2}$-P$_{1/2}$ transition, and a repumper laser at 866 nm, which addresses the D$_{3/2}$-P$_{1/2}$ transition (Fig. \ref{fig:EnergyDiagram}).  The relative isotope shifts for $^{44}$Ca$^{+}$ are $842$ MHz and $-4.5$ GHz for the S$_{1/2}$-P$_{1/2}$ and D$_{3/2}$-P$_{1/2}$ transitions, respectively~\cite{LucasPRA2004}.  

All lasers are commercially available. The photoionization lasers consist of a cavity-enhanced frequency-doubled external cavity diode laser (ECDL) at 423 nm (Toptica SHG 110) and a free-running 377 nm laser diode (Nichia).  The $^{40}$Ca$^+$ Doppler cooling lasers are a cavity-enhanced frequency-doubled ECDL with a tapered amplifier at 397 nm (Toptica TA-SHG 110) and an ECDL at 866 nm (Toptica DL100). The $^{44}$Ca$^+$ cooling/heating lasers are both ECDLs (Toptica DL100). Laser powers and beam profiles are measured using a Thorlabs S140A power meter (5$\%$ error) and a Thorlabs WM100-SI beam profiler (5$\%$ error). Laser intensities are converted to saturations, $s=I/I_s$,  where $I_s$ equals $4.7\times10^{-2}$ W/cm$^{2}$ for the 397-nm transition, and $3.4\times10^{-4}$ W/cm$^{2}$ for the 866-nm transition.  The lasers are pulsed using shutters with 2-ms response times (Uniblitz VS25). The laser frequencies are stabilized to a High Finesse WS7 wavemeter [10 MHz resolution and 60 MHz (3$\sigma$) absolute accuracy], with the absolute frequency calibrated to the trapped ion fluorescence.   

Ion fluorescence is collected simultaneously using an electron-multiplied CCD (EMCCD) camera (Princeton Instruments Photon Max 512) and a photon counter (Hamamatsu H7360-02 with a dark count rate of 50 counts/second).  A beam splitter directs 70$\%$ of the light to the photon counter which measures fluorescence during recooling.  The spatial resolution of the EMCCD is used to monitor ion position and potential ion loss. The collection efficiency at the photon counter (including all losses) is $10^{-4}$ of the photons scattered from the ion.    

\begin{figure}[htp]
\includegraphics[scale=0.25]{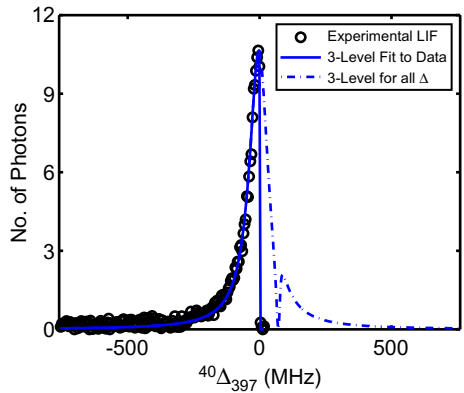}
\caption{\small (color online) Emission spectrum of the S$_{1/2}$-P$_{1/2}$ line of a calcium ion. Each data point (circles) is the number of photons acquired in 3 ms for each 397 nm laser detuning, $^{40}\Delta_{397}$ (averaged over 100 experiments). The solid blue line is a fit to the data using Eqn. (9) of Ref.~\cite{Gray1978} which theoretically describes the scattering rate for a three level system. The fitting function is evaluated over the range of $^{40}\Delta_{397}$ where fluorescence is observed. The blue dotted line shows the complete theoretical scattering profile across negative and positive detunings (assuming a motionless ion) when $^{40}\Delta_{866} = 20$ MHz, $^{40}s_{866}=1000$ and $^{40}s_{397}=8$.}\label{fig:397Scan3L}
\end{figure}

The absolute frequency calibration involves measuring the fluorescence spectra for each ion (Fig. \ref{fig:397Scan3L}) and fitting to a three-level system, which allows for coherent population trapping when $\Delta_{397}=\Delta_{866}$ (blue dot-dashed line) ~\cite{Gray1978}.  The only adjustable parameters are the frequency center of the S$_{1/2}$-P$_{1/2}$  and D$_{3/2}$-P$_{1/2}$ transitions. The resulting uncertainties in the absolute frequencies are  $13$ MHz for the S$_{1/2}$-P$_{1/2}$ transition and $20$ MHz in the D$_{3/2}$-P$_{1/2}$ transition.

\subsection{Experimental Procedure}

\begin{figure}[h]
\includegraphics[scale=0.4]{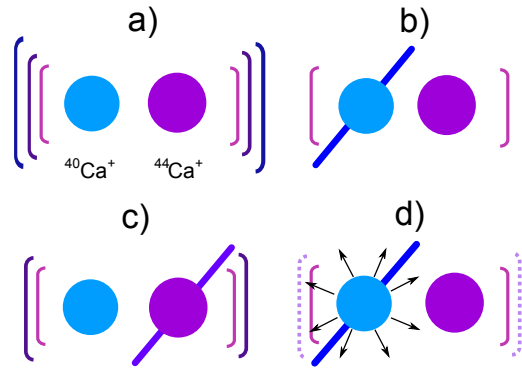}
\caption{\small (color online) Procedure for Sympathetic Heating Spectroscopy. The solid circles represent the two ions, the brackets represent the magnitude of vibrational energy in the Coulomb crystal, and the diagonal line represents an applied laser beam. (a) First, the spectroscopy ($^{44}$Ca$^{+}$) and control ($^{40}$Ca$^{+}$) ions are trapped. (b) The Coulomb crystal is then laser cooled via the control ion. (c) By simultaneously switching off the cooling laser and turning on the induced heating laser, the Coulomb crystal heats for a fixed interaction time ($t_{\rm heat}$). (d) Finally, the magnitude of heating is observed by measuring the fluorescence of the control ion as the crystal recools.}
\label{fig:SHS}
\end{figure}

Sympathetic Heating Spectroscopy (SHS) detects the scattering of photons from a spectroscopy ion by observing the heating and recooling of a control ion (Fig. \ref{fig:SHS}). Initially, the two ions are trapped (Fig. \ref{fig:SHS}(a)), and the spectroscopy ion is sympathetically cooled by a laser-cooled control ion (Fig. \ref{fig:SHS}(b)). By turning off the laser-cooling on the control ion and applying a near-resonant laser to the spectroscopy ion, the two-ion system will be heated (Fig. \ref{fig:SHS}(c)) for a time $t_{\rm heat}$. The resulting laser heating is measured by blocking the spectroscopy laser and monitoring the fluorescence of the control ion as it recools (Fig. \ref{fig:SHS}(d)).

This technique is demonstrated using two isotopes of Ca$^+$. In the experiments, the $^{40}$Ca$^+$ serves as the control ion and the $^{44}$Ca$^+$ as the spectroscopy ion. The $^{40}$Ca$^{+}$ lasers are detuned $^{40}\Delta_{866}= 20$ MHz and $^{40}\Delta_{397}= -30$ MHz from resonance with intensities fixed to yield saturation values of $^{40}s_{866}=1000$ and $^{40}s_{397}=8$. In the absence of ions in the trap, the photon counter reads $\sim$ 500 photons/second of background scattering at these laser intensities. The $^{44}$Ca$^{+}$ repumper laser has a fixed detuning $^{44}\Delta_{866}= 20$ MHz, while the $^{44}\Delta_{397}$ is varied to obtain spectra for a range of laser intensities.


\section{Results and Discussion}

\begin{figure}[htp]
\includegraphics[scale=0.4]{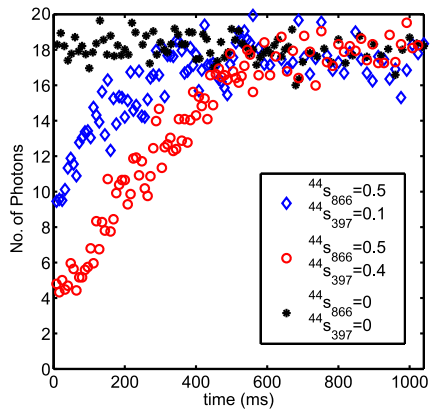}
\caption{\small (color online) Comparison of Doppler recooling fluorescence for three laser-induced heating parameters. In two heating situations, lasers are kept at $^{44}\Delta_{866} \approx 20$ MHz, $^{44}\Delta_{397} \approx 40$ MHz, and $^{44}s_{866}=0.5$. Blue diamonds and red circles show recooling after laser induced heating is applied on the spectroscopy ion with intensities proportional to $^{44}s_{397}=0.1$ and $^{44}s_{397}=0.4$, respectively, and for a period of $t_{\rm heat}=250$ ms. The black points show no deviation from steady-state fluorescence after turning off all lasers for the same period of time. Each time dependent curve is the average of 20 fluorescence trajectories. The ordinate shows the mean number of photons detected in 8 ms.} \label{fig:SHSHeatingCurve}
\end{figure}

Doppler recooling curves are obtained for a range of $t_{\rm heat}$, laser intensities, and detunings. Two representative curves are shown in Fig. \ref{fig:SHSHeatingCurve} illustrating the return of the $^{40}$Ca$^+$  fluorescence to steady state as the ions are cooled. In comparison, a third curve shows that without spectroscopy lasers there is no visible heating at $t_{{\rm heat}}=250$ ms. Each of the three curves is an average of 20 individual experiments and the data points measure the number of photons collected in a time $t_{\rm bin}$. Due to a limited number of available measurement bins (100 per experiment), the first 70 points are taken consecutively, and the last 30 points are taken every other time bin. This gives us the ability to observe the dynamic return to steady-state fluorescence at short times with a higher resolution. Depending on the experimentally observed fluorescence, $t_{\rm bin}$ is chosen to be between 3 and 8 ms to ensure that most of the trajectories reach a steady-state fluorescence. Recooling lasers remain on for an additional 500 ms ($t_{\rm int}$) to ensure that the system is initialized.

\begin{figure}[htp]
\includegraphics[scale=0.45]{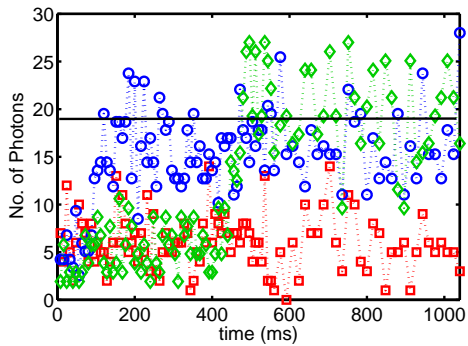}
\caption{\small (color online) Example of three single Doppler recooling fluorescence trajectories after heating of the Coulomb crystal for the average shown in Fig.~\ref{fig:SHSHeatingCurve} by the red circles. The black line represents the average steady state fluorescence over all 20 experiments. The ordinate shows the number of photons detected in 8 ms. Dotted lines are to guide the eye.} \label{fig:SHSHeatingCurve-slice}
\end{figure}

\begin{figure}
\includegraphics[scale=0.45]{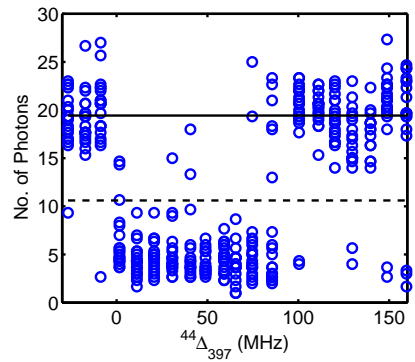}
\caption{\small(color online) Average of the first 3 fluorescence points for 20 Doppler recooling trajectories versus heating laser detunings. The ordinate shows the average number of photons detected in 8 ms. The solid black line shows the average steady-state fluorescence for a cold crystal. The dotted black line represents the threshold, which is 2$\sigma$ from the average of the steady-state fluorescence and is used to determine whether heating is observed during the experiment (spectroscopy laser parameters: $^{44}\Delta_{866} = 20$ MHz, $^{44}s_{866}=0.5$, and $^{44}s_{397}=0.4$).}
\label{fig:first3}
\end{figure}

The heating and recooling is stochastic, and the individual experiments show a variety of behaviors for the same laser parameters (Fig. \ref{fig:SHSHeatingCurve-slice}). To simplify the measurement, we report the percent of experiments where there is noticeable heating at short times, $P_{\rm heat}$. To calculate $P_{\rm heat}$, the average number of photons in the first three data points is compared to a threshold value, $T=\langle \Gamma_c\rangle t_{\rm bin}-2\sigma$, where $\langle \Gamma_c\rangle$ is the steady-state scattering rate and $\sigma$ is the standard deviation of the Poissonian distribution. If the signal is below the threshold, the experiment is marked as heating. False positives for cases without heating occur less than 2.5\% of the time. An example of applying this threshold to experimental data is shown in Fig. \ref{fig:first3}. A clear distinction between heating and no heating is observed.

SHS is compared to the expected fluorescence for a cold ion, 
\begin{equation}
I_{\rm LIF}=\langle\Gamma_{s}\rangle t_{ \rm meas}
\label{eqn:LIF}
\end{equation}
where $\Gamma_{s}$ is the steady state scattering rate of the spectroscopy ion ($^{44}$Ca$^+$) calculated using Eqn. (9) of Ref.~\cite{Gray1978}, and scaled by our experimental collection efficiency. In SHS, no signal is collected during $t_{\rm heat}$ and $t_{\rm int}$, but for LIF, photons can be collected the whole time, thus $t_{ \rm meas}=t_{\rm heat}+130\cdot t_{\rm bin} + t_{\rm int}$. In practice, many experiments should be averaged to obtain a fluorescence spectrum with a high signal to noise ratio.

An SHS signal is expected when $^{44}\Delta_{397}>0$, which is the opposite of the standard fluorescence spectra (Fig. \ref{fig:397Scan3L}) where fluorescence is detected when $^{44}\Delta_{397}<0$.  In the standard measurement, the laser cools the ion and reduces the Doppler shift to below the natural linewidth. As a result, the observed fluorescence can be described assuming the ion is motionless. For SHS, the ion is heated by the scattered photons and the resulting Doppler shift dramatically changes the scattering rate. How the total heating depends on the interaction time  and the laser intensities is difficult to calculate.  Experiments examining the limits of detecting this heating are now described.

\subsection{Effect of laser powers on SHS}

\begin{figure}
\includegraphics[scale=0.5]{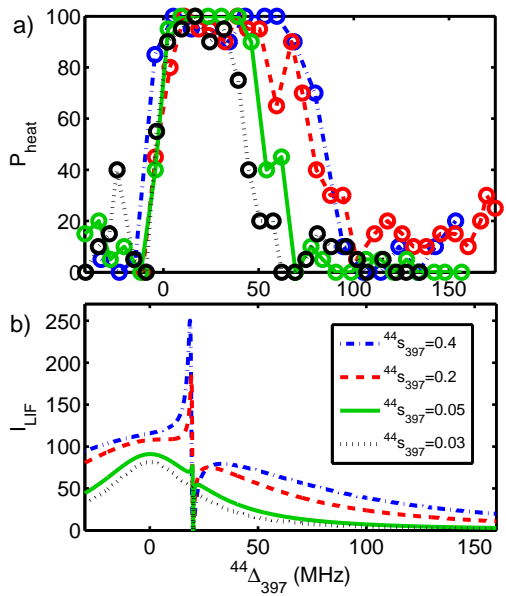}
\caption{\small(color online) (a) The effect of varying the power of the heating laser, $^{44}s_{397}$, on the SHS spectra. The repumper laser is held constant at $^{44}\Delta_{866} = 20$ MHz with an intensity proportional to $^{44}s_{866}=0.5$. Percentages are based on 20 experiments for each $^{44}\Delta_{397}$ with $t_{\rm bin}=8$ ms and $t_{\rm heat}$=250 ms. SHS peaks are clearly saturated, and broadening is correlated with the intensity of the 397 heating laser. Lines connecting experimental data are to guide the eye. (b) Simulated $I_{\rm LIF}$ using the same experimental parameters and $t_{\rm meas}=$ 1.79 s. A monotonic broadening of the profile is seen with increasing $^{44}s_{397}$ at the chosen set of repumper experimental parameters.}
\label{fig:SHSLIFBlue}
\end{figure}

The photon scattering rate and laser heating are most strongly affected by the power of the 397 nm laser. Fig. \ref{fig:SHSLIFBlue}(a) shows the variation in the SHS spectra with 397 nm laser power.  The spectra has a wide range of detunings yielding equivalent signals. The maximum signal is detected with $^{44}\Delta_{397}$ between $+10$ and $+70$ MHz. The spectra have a sharp rise at zero detuning, and the width of the peak reduces with laser power resembling the trend observed for the calculated LIF spectra in Fig. \ref{fig:SHSLIFBlue}(b).  For certain laser
intensities and detunings, the three-level system shows fine features
arising from both resonance with dressed states and coherent
population trapping \cite{Armindo1996}. Following Ref.\cite{Gray1978}, the
calculations assume lasers without linewidths and a motionless ion.     

\begin{figure}
\includegraphics[scale=0.5]{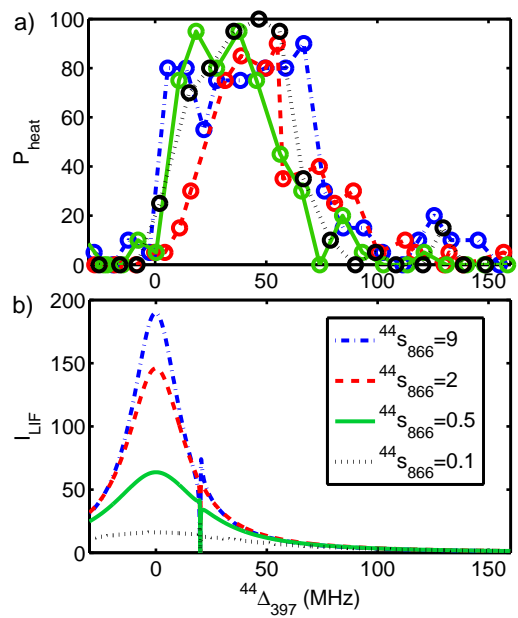}
\caption{\small(color online) The effect of varying the power of the heating repumper laser, $^{44}s_{866}$, at constant detuning ($^{44}\Delta_{866} = 20$ MHz), on the SHS spectra.  Percentages are based on 20 experiments for each $^{44}\Delta_{397}$, which is held at a constant $^{44}s_{397}$ of $0.03$ ($t_{\rm bin}=5$ ms and $t_{\rm heat}=250$ ms). Lines connecting experimental data are to guide the eye. (b) Simulated $I_{\rm LIF}$ using the same experimental parameters and $t_{\rm meas}=$ 1.4 s. The intensity of the repumper laser does not appear to have an effect on the linewidth when using a very low 397 saturation.}
\label{fig:SHSLIFIR}
\end{figure}

A difficulty in the measurement of atomic and molecular spectra by fluorescence arises from the existence of metastable states. In Ca$^+$, the effective lifetime of the metastable D$_{3/2}$ state can be controlled by the intensity of the 866 nm laser. Example SHS spectra are shown in Fig. \ref{fig:SHSLIFIR}(a) for varying $^{44}s_{866}$.  A less pronounced effect on peak width is observed compared to changing $^{44}s_{397}$, which is the same trend observed in the predicted LIF spectra shown in Fig.~\ref{fig:SHSLIFIR}(b). SHS spectra show laser induced heating more than 50\% of the time at $^{44}\Delta_{397}$ between $+5$ and $+65$ MHz for all of the experimental repumper laser intensities. 

\subsection{Effect of $t_{\rm heat}$ on SHS}

\begin{figure}[htp]
\includegraphics[scale=0.5]{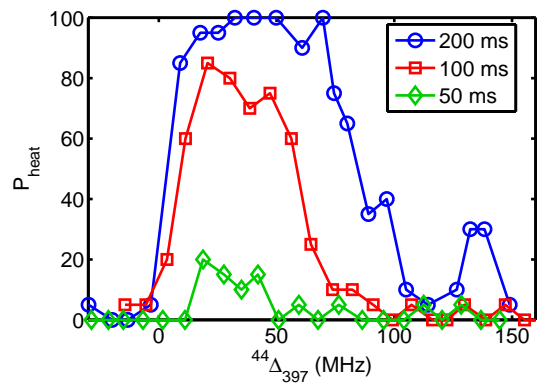}
\caption{\small (color online) The effect of varying $t_{\rm heat}$ on SHS spectra. The intensity of the 397 spectroscopy laser is proportional to $^{44}s_{397}=0.03$, and the heating repumper laser is tuned to $^{44}\Delta_{866} = 25$ MHz with a saturation value fixed at $^{44}s_{866}=0.5$. Lines shown are to guide the eye.}
\label{fig:SHSHeatingTime}
\end{figure}

In order to determine the limits of observable heating, the laser powers were decreased and the heating time extended. The heating time plays a large role in the accumulative heating mechanism, as shown in Fig. \ref{fig:SHSHeatingTime}. The $P_{\rm heat}$ line profiles not only increase in height, but also broaden. For long heating times the red edge of the peak is within 10 MHz of the S$_{1/2}$-P$_{1/2}$ transition (blue circles and red squares). This is not the case as the heating time is reduced (green diamonds). In future work, we plan to connect the peak of the accumulated heating spectra to the transition peak center and linewidth.

\begin{figure}[htp]
\includegraphics[scale=0.5]{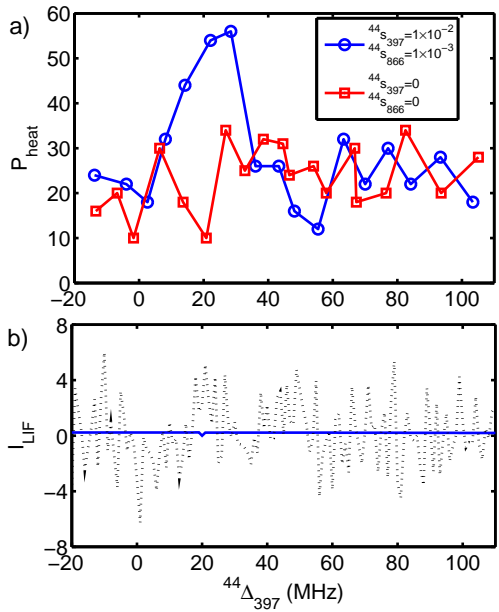}
\caption{\small (color online) Comparison of a low power SHS spectra with simulated LIF signal. (a) The blue circles show a SHS spectra for $^{44}s_{397}$=0.01 and $^{44}s_{866}$=1$\times 10^{-3}$ with $t_{\rm heat}$= 1 s, $t_{\rm bin}= 3$ ms, and $^{44}\Delta_{866}=20$ MHz (averaged over 20 experiments). The background $P_{\rm heat}$ measurements without heating lasers are shown by the red squares. Lines are to guide the eye. (b) The predicted LIF spectrum with a $t_{\rm meas}=1.89$ s is shown in blue for comparison. The dotted black line is a simulation of the shot noise corresponding to the dark counts of the experimental device (50 counts/sec).}
\label{fig:SHSLIFlimit}
\end{figure}

The heating rate of the trap limits the amount of laser induced heat that can be detected. For a single ion, no heating is measured up to 20 seconds. For two ions, a small amount of heating is observable at 1 second corresponding to a $P_{\rm heat} \approx 10 - 30$ and a temperature of approximately 7 K. Fig.~\ref{fig:SHSLIFlimit}(a) shows that for very low laser powers ($^{44}s_{397}$=0.01 and $^{44}s_{866}=1\times10^{-3}$), the $P_{\rm heat}$ spectrum has a signal to noise ratio of $\sim$ 2. The predicted LIF spectrum at these conditions is shown in Fig.~\ref{fig:SHSLIFlimit}(b), together with a simulation of the photon counting noise associated with the dark counts of the device employed in these experiments. The number of scattered photons ($I_{\rm LIF}$) is at least 9 times smaller than the photon shot noise. For $t_{\rm meas}=2$ s and averaging over 20 experiments, an $I_{\rm LIF}>4.5$ would be required to distinguish the laser induced fluorescence from noise in the detector with high confidence. One way to overcome the shot noise is to average over more than 7000 experiments. Alternatively, an increase in the collection efficiency will reduce the time required to obtain a LIF spectra. For SHS, the same detection improvement would enhance our ability to distinguish laser induced heating from trap heating.

A SHS signal is observed when the calculated optimal fluorescence would result in at most 1500 photons per second being scattered into a solid angle of 4$\pi$. This is very dim compared to the millions of photons per second typically scattered by alkaline earth ions. Experiments with extremely low heating powers, corresponding to a maximum fluorescence of a few hundred photons from the spectroscopy ion, resulted in indistinguishable signal from the case without heating lasers.

\section{Conclusion}
\indent Sympathetic heating spectroscopy is a sensitive way to detect spectral lines in ions. In the experiment, a few scattered photons from the spectroscopy ion are transformed into a large deviation from steady-state fluorescence on the control ion. Although application to low-scattering rate transitions is natural, this technique would be most useful for transitions at the frequency limit of detectors.

\indent  The current work uses a simple metric, $P_{\rm heat}$, to measure the heating in a two ion Coulomb crystal. This reveals the approximate line position but does not provide a clear method for determining the natural linewidth. The resolution of the spectrum is limited by the accumulative stochastic heating mechanism.  A better understanding of the distribution of ion energies after heating and during the recooling process may allow for the extraction of the transition dipole moment. Future work will include developing a Monte Carlo simulation that accounts for the effect of stochastic scattering of photons on the ion motion.

\indent Furthermore, this work suggests a middle ground between QLS and SHS where the atomic sidebands are used to determine the temperature. This method will allow for the detection of even lower scattering rates with the limit being the absorption of a single photon as demonstrated by QLS~\cite{SchmidtScience2005}. Either this intermediate technique, or perhaps an improved SHS, would be able to resolve the peak centers of the Fe$^+$ lines between 234-260 nm for comparison with high-redshift astronomical data~\cite{WebbPRL1999}. An immediate improvement to SHS can be obtained by simply increasing the collection efficiency.  

Based on these observations for a two ion Coulomb crystal, it is possible to detect the heat induced by less than 1500 scattered photons. Recent results showing significant photon scattering from SrF using two lasers to address the (0,0) and (0,1) vibrational bands of an electronic transition, together with sidebands from an electro-optic modulator (EOM) to address hyperfine structure ~\cite{ShumanPRL2009}, suggest that a clever choice of molecular ion can lead to similar rates. We envision an experiment in which a control loop optimizes intensities, laser frequency, and the position of EOM sidebands to maximize the radiative force that heats the Coulomb crystal. The parameter results obtained by the control loop will encode information about the internal structure of the molecular ion.

\begin{acknowledgments}
The authors thank Prof. Shinji Urabe of Osaka University for the ion trap and Dr. Richart Slusher for the loan of a 397 nm laser. This work was supported by the Georgia Institute of Technology and by IARPA through the Army Research Office award W911NF-08-1-0515. 
\end{acknowledgments}


%

\end{document}